
\magnification 1200  \baselineskip=13pt  \hfuzz=5pt \null
\overfullrule=0mm
\def\sq{\hbox {\rlap{$\sqcap$}$\sqcup$}}
\def\R{ {\rm R \kern -.31cm I \kern .15cm}}
\def\C{ {\rm C \kern -.15cm \vrule width.5pt \kern .12cm}}
\def\Z{ {\rm Z \kern -.27cm \angle \kern .02cm}}
\def\N{ {\rm N \kern -.26cm \vrule width.4pt \kern .10cm}}
\def\1{{\rm 1\mskip-4.5mu l} }
{\nopagenumbers
\line{\hfil CERN-TH/95-152}
\line{\hfil LPTHE Orsay 95/32}
\line{\hfil ENSLAPP-A558-95}
\line{\hfil Revised Version}
\vfil
\centerline{{\bf GENERALIZATION OF THE CALOGERO-COHN BOUND
}} \smallskip  \centerline{{\bf ON THE NUMBER OF BOUND STATES}}
\bigskip

\centerline {\bf K. Chadan} \par
\centerline{Laboratoire de Physique Th\'eorique et Hautes
Energies\footnote{*} { Laboratoire associ\'e au Centre
National de la Recherche Scientifique URA D0063.}}
\centerline{B\^atiment 211, Universit\'e de Paris-Sud,
91405 Orsay cedex}

\vskip 1.3 mm
\centerline{\bf R. Kobayashi} \centerline{Department of Mathematics,
Science
 University of
Tokyo}
\centerline{Noda, Chiba 278, Japan}

\vskip 1.3 mm
\centerline{\bf A. Martin}
\centerline{Theory Division, CERN, 1211 Geneva 23, Switzerland}
\centerline{and}
\centerline{Laboratoire de Physique Th\'eorique
ENSLAPP\footnote{**}{URA 14-36 du CNRS associ\'ee \`a L'Ecole
Normale Sup\'erieure de Lyon et \`a l'Universit\'e de Savoie.}}
\centerline{Groupe d'Annecy, LAPP, B.P. 110}
\centerline{F-74941 Annecy-le-Vieux, France}

\vskip 1.3 mm
\centerline{and}
\vskip 1.3 mm
\centerline{\bf J. Stubbe}
\centerline{Department of Theoretical Physics, University of Geneva}
\centerline{4 Quai Ernest-Ansermet, 1221 Geneva 4, Switzerland}
\vfil
\noindent {\bf \underbar{Abstract}} \par
It is shown that for the Calogero-Cohn type upper bounds on the number
of bound
 states of a
negative spherically symmetric potential $V(r)$, in each angular
momentum state,
 that is,
bounds containing only the integral $\int_0^{\infty} |V(r)|^{1/2} dr$,
the
 condition $V'(r)
\geq 0$ is not necessary, and can be replaced by the less stringent
condition
 $(d/dr)
[r^{1-2p}(-V)^{1-p}] \leq 0$, $1/2 \leq p < 1$, which allows
oscillations in the
 potential.
The constants in the bounds are accordingly modified, depend on $p$ and
 $\ell$, and tend to the standard value for $p = 1/2$.
\vfil
\noindent{\obeylines CERN-TH/95-152
\noindent LPTHE Orsay 95/32
\noindent ENSLAPP-A558-95
\noindent October 1995}
\eject }
\pageno=1
\noindent {\bf I - \underbar{Introduction}} \vskip 5 mm
Among the numerous bounds on the number of bound states in a potential,
or
more generally the moments of the eigenvalues, physicists prefer those
given
by semi-classical expressions. Often these are only valid in the strong
coupling limit and the price to pay to convert them into strict bounds
is to
multiply them by some appropriate numerical factor. In the case of the
bound
of Calogero and Cohn, who assume monotonicity of the potential, it is a
factor two. What we shall show in this paper is that the requirement of
monotonicity of the potential can be considerably weakened if one is
ready to
replace this factor two by a correspondingly larger one. This will
broaden
the field of application of the bound.

 For a regular and spherically symmetric potential $V(r)$, which
is purely
 attractive and
nondecreasing $(V' \geq 0)$, and vanishes at infinity, Calogero$^1$ and
Cohn$^2$
 have shown
that in the $S$-wave, the number of bound states for the radial
Schr\"odinger
 equation
$$\varphi '' + E \varphi = V \varphi \ \ , \quad r \in [0, \infty) \ \ ,
\quad
 \varphi (0)
= 0 \eqno(1)$$

\noindent admits the upper bound
$$n_0 \leq {2 \over \pi} \int_0^{\infty} |V(r)|^{1/2} dr \ \ \ .
\eqno(2)$$

\noindent By regular potentials, we mean those which are less singular
than
 $r^{-2}$ at the
origin, and go to zero faster than $r^{-2}$ at infinity. More precisely,
they
 satisfy the
condition$^{3,4}$
$$\int_O^{\infty} r |V(r) | dr < \infty \ \ \ . \eqno(3)$$

\noindent In what follows, we always assume that this condition is
satisfied.
 \par

Recently, the bound has been generalized to higher angular momenta by
taking
 into account
the effect of the centrifugal potential $\ell (\ell + 1)/r^2$, $\ell
\geq 0$.
 One finds
then, again, with the same conditions on the potential, namely (3) and
$V' \geq
 0$, the
upper bound$^5$
$$n_{\ell} \leq 1 + {2 \over \pi} \left [ \int_0^{\infty} |V|^{1/2} dr -
 \sqrt{\left ( {\pi
\over 2} \right )^2 + \ell (\ell + 1)} \right ] \eqno(4)$$

\noindent which reduces to (2) for $\ell = 0$. \par

For negative values of $\ell$, $- 1/2 < \ell \leq 0$, and again with the
 previous
conditions on $V$, one has now$^5$
$$n_{\ell} \leq {1 \over \sqrt{2(2 \ell + 1)}} \int_0^{\infty} |V|^{1/2}
dr \ \
 \ .
\eqno(5)$$

\noindent Making $\ell = 0$ here, we do not get the Calogero-Cohn
constant
 $2/\pi$, but
$\sqrt{2}/2$, which is slightly larger. The above bound is singular for
$\ell =
 - 1/2$,
and we shall see that this cannot be avoided. \par

With no condition on the potential, except (3), we have the general
Bargmann
 bound$^{3,6}$
$$n_{\ell} \leq {1 \over (2 \ell + 1)} \int_0^{\infty} r |V(r)|dr \ \ \
 \eqno(6)$$

\noindent This bound also has been generalized, again with no condition
on
 potential
except (3), to a large family  of bounds$^7$
$$n_{\ell} \leq {C_p \over (2 \ell + 1)^{2p-1}} \int_0^{\infty} |r^2V|^p
{dr
 \over r}
\eqno(7)$$

\noindent where $p$ is a free parameter, $1 \leq p \leq 3/2$, and
$$C_p = {(p - 1)^{p-1} \Gamma (2p) \over p^p \Gamma^2(p)} \ \ \ .
\eqno(8)$$

\noindent Making $p \downarrow 1$, we get the Bargmann bound (6), as
expected.
 In (6) and
(7), we have tacitly assumed $V$ to be negative everywhere. If the
potential
 changes
sign, then we should replace in (6) and (7) $V$ by its negative part
$V_-$. \par

Now, the question arises whether one could fill the gap between (7),
valid for
 $1 \leq p
\leq 3/2$, and (2), (4), and (5), where we have the integral of
$|V(r)|^p$, with
 $p =
1/2$. In short, whether one could find, with some condition on the
potential -
 similar
to the Calogero-Cohn condition $V' \geq 0$ - such that one would have a
bound
 similar to
(7) for $1/2 \leq p \leq 1$. The answer is in the affirmative. Indeed,
assuming
 again
$V$ to be negative everywhere, and
$${d \over dr} \left [ r^{1-2p}(-V)^{1-p} \right ) \leq 0 \ \ , \quad {1
\over
 2} \leq p
\leq 1 \ \ \ , \eqno(9)$$

\noindent one can show that$^8$
$$n_{\ell} \leq {p \over (1 - p)^{1-p} (2 \ell + 1)^{2p-1}}
\int_0^{\infty}
 \left ( - r^2 V
\right )^p {dr \over r} \ \ \ . \eqno(10)$$

We should remark here that again, for $p = 1$, (9) imposes no condition
on the
 potential,
and (10) gives us then the Bargmann bound (6), as expected. On the other
hand,
 for $p =
1/2$, we obtain the Calogero-Cohn condition $V' \geq 0$, but then (10)
goes to
 (5) with
$\ell = 0$, which is slightly larger than (2), as we have noticed
before. \par

For $p$ strictly inside the interval $(1/2, 1)$, the potential may have
 oscillations while
staying everywhere negative. As examples, we give just the two following
 ones$^9$~:
$$V_1 = - r^{(2p - 1)/(1 - p)} \ e^{-r/(1 - p)} \ \ \ ; \eqno(11)$$
$$V_2 = - r^{(2p - 1)/(1 - p)} \left \{ \left [ 1 + {1 \over 2} (\sin r
+ \cos
 r)
\right ] e^{-r} \right \}^{1/(1 - p)} \ \ \ . \eqno(12)$$

\noindent It is easily seen that $V_1$, which vanishes at the origin,
has a
 minimum before
going to zero at $r = \infty$, whereas $V_2$ oscillates indefinitely
while going
 to zero at
$r = \infty$. Both satisfy (9). \par

The purpose of the present paper is to show that, in fact, condition (9)
leads
 to a
Calogero-Cohn type bound, for all $p \in [1/2, 1)$, that is, a bound
containing
 the
integral $\int_0^{\infty} \sqrt{|V|} dr$, but, of course, with a
different
 constant than
$2/\pi$. This would be much more satsifactory for strong attractive
potentials
 since we
know that, in the limit $\lambda \to \infty$, the number of bound states
of
 $\lambda V$,
for any fixed $\ell \geq 0$, has the asymptotic behaviour$^{10,11}$
$$n_{\ell} = {\lambda^{1/2} \over \pi} \int_0^{\infty} |V_-|^{1/2} dr +
 \hbox{\rm smaller
terms} \ \ \ , \eqno(13)$$

\noindent with no condition on $V$ than the finiteness of the integral.
\vskip 5 mm
\noindent {\bf II - \underbar{General proof of a Calogero-Cohn type
bound}}
 \vskip 5 mm
Since we assume in general (3) and (9), we must have $\lim r^2 V(r) = 0$
as $r
\to 0$ or
 $r \to
\infty$. Now, (9) can be written
$${d \over dr} \left \{ \left [ r^2(-V) \right ]^{1-p}/r \right \} = -
q(r)
 \eqno(14)$$

\noindent where $q$ is some positive function, and the function inside
the
 bracket in the
l.h.s. vanishes at infinity. Assuming $q(r)$ to be integrable there,
which is
 quite
natural, and solving the differential equation (14) for $V$, together
with
 $V(\infty ) =
0$, we obtain
$$V(r) = - r^{{2p - 1 \over 1 - p}} \left ( \int_r^{\infty} q(t) dt
\right )^{{1
 \over 1
- p}} \ \ \ . \eqno(15)$$

\noindent The only condition to be imposed on $q(r)$, besides being
positive and
 integrable
for $r > 0$, is that (3) must be satisfied. As examples, we can take
$q(t) =
 e^{-t}$ or
$q(t) = (1 + \sin t)e^{-t}$, and we obtain, respectively, (11) and (12).
\par

We have now to deal with the radial Schr\"odinger at zero energy
$$\varphi ''_{\ell} = \left [ V(r) + {\ell (\ell + 1) \over r^2} \right ]
 \varphi_{\ell}
\eqno(16)$$

\noindent together with $\varphi_{\ell}(0) = 0$, and the well-known
nodal
 theorem$^3$,
which asserts that the number of bound states $n_{\ell}(V)$ is equal to
the
 number of nodes
(zeros) of $\varphi_{\ell}(r)$ on the real $r$-axis, origin excepted.
The
 potential being
given by (15), we can now use the Liouville transformation
$$r \to Z = r^{1/2(1 - p)} \ \ , \quad \varphi \to \psi (Z) = Z^{(2p -
1)/2}
 \varphi (r(Z))
\ \ \ . \eqno(17)$$

The change of variable is one-to-one, and applies $r \in [0, \infty )$
on $Z \in
 [0, \infty
)$, $r = 0$ corresponding to $Z = 0$. The change of function is such
that to
 $\varphi (r =
0) = 0$ corresponds $\psi (Z = 0) = 0$. After the transformation, (16)
becomes
 $(\dot{\ } =
d/dZ)$
$$\ddot{\psi}(Z) = \left [ \widetilde{V}(Z) + {L(L + 1) \over Z^2}
\right ] \psi
 (Z)
\eqno(18)$$

\noindent together with $\psi (0) = 0$, where
$$\left . \widetilde{V}(Z) = - 4(1 - p)^2 \left ( \int_r^{\infty} q(t)
dt \right
 )^{{1 \over
1 - p}} \right |_{r=Z^{2(1 - p)}} \ \ \ , \eqno(19)$$

\noindent and
$$L = L(\ell , p) = - {1 \over 2} + (1 - p) (2 \ell + 1) \ \ \ .
\eqno(20)$$

We see that we have again to deal with a Schr\"odinger equation at zero
energy,
 in the
variable $Z$, with Dirichlet condition at the origin, and a potential
 $\widetilde{V}(Z)$
which is now attractive and increasing, together with the centrifugal
term $L(L
 + 1)/Z^2$.
Moreover, the zeros of $\psi$ on the positive $Z$-axis are in one-to-one
 correspondence
with those of $\varphi (r)$ on the positive $r$-axis. It follows that
the number
 of bound
states is the same for (16) as for (18). However, the advantage of (18)
is that
 we can
now use the bounds (4) or (5), according to the value of $L \geq 0$, or
$- {1
 \over 2} <
L \leq 0$. We would then get bounds which contain $\int_0^{\infty}
|\widetilde{V}(Z)|^{1/2} dZ$. This, expressed in terms of the variable
$r$, is
 exactly
$\int_0^{\infty} |V(r)|^{1/2} dr$, and we obtain the desired result. In
 r\'esum\'e, we have
the following \par \vskip 5 mm
\noindent \underbar{\it Theorem 1} \par
Under the condition (9) on the potential, we have
$$n_{\ell} \leq 1 + {2 \over \pi} \left [ \int_0^{\infty} |V(r)|^{1/2}
dr -
 \sqrt{\left (
{\pi \over 2} \right )^2 + L(L + 1)} \right ] \eqno(21)$$

\noindent if $L$, given by (20), is $\geq 0$, and
$$n_{\ell} \leq {1 \over \sqrt{2(2L + 1)}} \int_0^{\infty} |V(r)|^{1/2}
dr
 \eqno(22)$$

\noindent if $L \in ( - {1 \over 2} , 0]$. The first case corresponds to
$\ell
 \geq (2p
- 1)/4(1 - p)$, and the second to $- 1/2 < \ell \leq (2p - 1)/4(1 - p)$.
We
 should
remark here that when $\ell = 0$, we have $L = {1 \over 2} - p$, which
is
 negative,
except for $p = {1 \over 2}$, and so, we must use in general (22). If
$\ell =
 0$, and $p
= {1 \over 2}$, i.e. the Calogero-Cohn case, we have the bound (2), and
don't
 have to use
(22).
\vskip 5 mm
\vfill\eject
\noindent {\bf III - \underbar{Direct proof of the generalized bound
(22)}}
\vskip 5 mm
We consider the $S$-wave, described by equation (1). Suppose that there
are
 $n_0$ bound
states. This means that $\varphi_0(r)$ has $n_0$ nodes $0 < r_1 < \cdots
 r_{n_0} < \infty$.
We assume now the potential to satisfy the following conditions~:
$$V(r) \leq 0 \ \ , \quad \left [ - r^{- \nu} V(r) \right ] ' \leq 0
\quad
 \hbox{\rm for
some} \ \nu \geq 0 \ \ \ . \eqno(23)$$

\noindent We have now~:
\vskip 5 mm
\noindent \underbar{\it Lemma 1} \par
If $V$ satisfies (23), the same is true for $- V(r) (r - r_k)^{- \nu}$
for $r >
 r_k$. Indeed,
we have
$$|V(r)| \left ( r - r_k \right )^{- \nu} = \left [ |V(r)| r^{- \nu}
\right ]
 \left ( {r
\over r - r_k} \right )^{\nu} \ \ \ ,$$

\noindent and both factors are decreasing for $r \geq r_k$. $\sq$ \par

Now, from the Bargmann bound (6) with $\ell = 0$, we have
$$1 \leq \int_{r_k}^{r_{k+1}} \left ( r - r_k \right ) |V(r)| dr \ \ \ .
$$

\noindent Taking $\rho_k = r - r_k$, this can be written
$$1 \leq \int_0^{r_{k+1}-r_k} \rho_k W(\rho_k) d\rho_k \eqno(24)$$

\noindent where $W(\rho_k) = V(\rho_k + r_k)$. Restricting ourselves to
the
 interval
$(r_k, r_{k+1})$, and dropping the index $k$, let us define
$$I(\rho ) = \int_0^{\rho} \sqrt{|W( \rho ')|} d\rho ' \ \ \
.\eqno(25)$$

\noindent If $V$ satisfies (23), Lemma 1 shows that $|W( \rho )| \rho^{-
\nu}$
 is also
decreasing. Therefore
$$I(\rho ) = \int_0^{\rho} \sqrt{|W( \rho ') | \rho '^{- \nu}} \rho
'^{\nu/2}
 d\rho ' \geq
\sqrt{\rho^{- \nu} |W(\rho )|} \times {\rho^{1 + {\nu \over 2}} \over 1
+ {\nu
 \over 2}} =
\rho {\sqrt{|W( \rho )|} \over 1 + {\nu \over 2}} \ \ \  . \eqno(26)$$

Using now this inequality, together with (24) and ${dI \over d \rho} =
 \sqrt{|W(\rho )|}$,
we obtain
$$1 \leq \int_0^{r_{k+1} - r_k} \left [ \rho \sqrt{|W(\rho )|} \right ]
 \sqrt{|W( \rho )|}
d \rho \leq \left ( 1 + {\nu \over 2} \right ) \int_0^{r_{k+1} - r_k}
I(\rho )
 {dI \over d
\rho } d \rho =$$ $${1 \over 2} \left ( 1 + {\nu \over 2} \right ) \left
[ I
 \left ( r_{k+1}
- r_k \right ) \right ]^2  \ \ \ .$$

\noindent Therefore
$$1 \leq {\sqrt{\nu + 2} \over 2} I \left ( r_{k+1} - r_k \right ) =
{\sqrt{\nu
 + 2} \over
2} \int_{r_k}^{r_{k+1}} \sqrt{|V(r)|} dr \ \ \ . \eqno(27)$$

\noindent Adding up these inequalities for all the intervals, we end up
with
$$n_0 \leq {\sqrt{\nu + 2} \over 2} \int_0^{\infty} |V|^{1/2} dr \ \ \ .
 \eqno(28)$$

In order to apply this inequality to our potential satisfying (9), we
just have
 to put $\nu
= (2p - 1)/(1 - p)$. When $p$ varies between ${1 \over 2}$ and 1, $\nu$
varies
 between 0
and $\infty$. In any case, we obtain finally
$$n_0 \leq {1 \over 2 \sqrt{1 - p}} \int_0^{\infty} |V(r)|^{1/2} dr
\eqno(29)$$

\noindent which is the desired result. $\sq$ \par

Let us remark here that, for $p = 1$, the r.h.s. of (29) is infinite.
Indeed,
 condition (9)
for $p = 1$ puts no restriction on the potential. And with no
restriction on the
 potential,
the only bound which is valid in general is the Bargmann bound (6),
which
 contains the
integral of $|V|$ instead of $|V|^{1/2}$. As is well-known$^{3,6}$, the
Bargmann
 bound can
be saturated by $n_0$ negative $\delta$-potentials with suitable
strengths and
 locations~:
$$V(r) = - \sum_{k=1}^{n_0} g_k \delta \left ( r - r_k \right )
\eqno(30)$$

\noindent and such a potential gives zero in the r.h.s. of (29). It
follows that
 the
singularity at $p = 1$ in front of the integral in (29) cannot be
avoided. We
 can
summarize the results in the following
\vskip 5 mm
\noindent \underbar{\it Theorem 2} \par
For a purely attractive potential satisfying the condition (23) for some
$\nu
 \geq 0$,
the number of $S$-wave bound states satisfies the bound
$$n_0 \leq {\sqrt{\nu + 2} \over 2} \int_0^{\infty} |V|^{1/2} dr \ \ \
 .\eqno(31)$$

\vskip 5 mm
\noindent {\bf IV - \underbar{Improvement of the Generalized Bound
(22)}}
\vskip 5 mm

If $L = L(\ell,p) \in (-1/2,0)$ we have the following operator
inequality due to the local
uncertainty principle$^{12}$
$$
-{d^2 \over dz^2} + {L(L+1) \over z^2} \geq - (2L+1)^2 {d^2 \over
dz^2}~.
\eqno(32)
$$
Hence the number of bound states of the operator associated to (18) is
bounded above by the number
of bound states of $(-2L+1)^2{d^2 \over dz^2} + \tilde V(z)$.  Applying
the Calogero-Cohn bound to
this operator we find
$$
n_\ell \leq {1 \over 2L+1} {2 \over \pi} \int^\infty_0 |V(z)||^{1/2}
dr~.
\eqno(33)
$$
Together with (22) we therefore have
\par \vskip 5 mm
\noindent \underbar{\it Theorem 3} \par
Under the condition (9) on the potential and if $L \in (-1/2,0]$, we
have
$$
n_\ell \leq C_L \int^\infty_0 |V(r)|^{1/2} dr
\eqno(34)
$$
where
$$
C_L = {\rm min} \left( {1 \over 2L+1} {2 \over \pi}
, {1 \over \sqrt{2(2L+1)}} \right)~.
\eqno(35)
$$
If the potential satisfies condition (23) for some $V \leq 0$, the
number of $S$-wave bound states
satisfies the bound
$$
n_0 \leq C_\nu \int^\infty_0 |V(r)|^{1/2} dr
\eqno(36)
$$
with
$$
C_\nu = {\rm min} \left( {\nu+2 \over \pi},{\sqrt{\nu+2} \over 2}
\right)
\eqno(37)
$$
In particular
$$
C_\nu =
\pmatrix{
{\nu+2 \over \pi} & {\rm if} & \nu < \nu_C:= {\pi^2-8 \over 4} \cr
{\sqrt{\nu+2} \over 2} & {\rm if} & \nu_C \leq \nu < \infty}
\eqno(38)
$$
We note that $\lim_{\nu\rightarrow 0} C_\nu = 2/\pi$
 so that we recover in the limit $\nu \rightarrow
0$ the optimal bound.

\vskip 5 mm
\noindent {\bf V - \underbar{The case $0 > \nu > - 2$}}
\vskip 5 mm

{}From (31), we see that if the r.h.s. is less than 1, then there is no
bound
 state. It is
easy to see that this holds not only for $\nu > 0$, but also for $0 >
\nu > -
 2$. In other
words, if the absolute value of the potential decreases faster than
 $r^{-|\nu|}$, $0 > \nu
> - 2$, then the condition
$${\sqrt{\nu + 2} \over 2} \int_0^{\infty} |V(r)|^{1/2} dr < 1
 \eqno(39)$$

\noindent guarantees the absence of bound states. As an example, we can
consider
 the Yukawa
potential $V = - g \exp (- \mu r)/r$. Here, $\nu = - 1$, and the
constant in
 front of the
integral becomes 1/2, which is smaller than the Calogero constant
$2/\pi$.
 However, we must
remember that all this is derived from the Bargmann bound, and therefore
we
 cannot do better
than that. Also, generalization to $n$ bound states for $\nu < 0$ is
impossible
 because
Lemma 1 no longer holds. So the interest of $\nu < 0$ is rather limited.

\vskip 5 mm
\noindent {\bf VI - \underbar{The Calogero's sufficient condition}}
\vskip 5 mm

Here, we would like to see how good is the constant $C_\nu$ in front
 of (36).
For this purpose, we consider the simple power-potential
$$V(r) =  - \lambda r^{\nu} \theta (1 - r) \ \ \ . \eqno(40)$$

\noindent Now, a sufficient condition of Calogero$^{13}$ states that,
for a
 purely
attractive potential, if there exists an $R$ such that
$${1 \over R} \int_0^{R} r^2 |V(r)| dr + R \int_R^{\infty} |V(r)| dr
\geq 1 \ \
 \ ,
\eqno(41)$$

\noindent then there is at least one bound state. $R$ is arbitrary here,
and can
 be chosen
at will. Applying (41) to (40), and maximizing with respect to $R$, we
find that
 it is sufficient to have
$$\lambda \geq \lambda_1 = (\nu + 2) \left [ {2 ( \nu + 2) \over \nu +
3} \right
 ]^{{1
\over \nu + 1}} \ \ \ . \eqno(42)$$

Now, assume that the r.h.s. of (36) is too large, and that $C_\nu$
 can be
replaced by a better (smaller) constant $\tilde C_{\nu}$. Applying now
(36), with $n_0
 = 1$ and
the better constant $\tilde C_{\nu}$ to our potential (40), we find that
we must have
 $\tilde C_{\nu}
\sqrt{\lambda_1} \int_0^1 r^{\nu/2} dr \geq 1$, that is
 $$\tilde C_{\nu} \geq {(\nu +
 2) \over 2
\sqrt{\lambda_1}} > \left ( {\sqrt{\nu + 2} \over 2}
\right ) \left [ {\nu + 3 \over 2 (\nu + 2)} \right ]^{{1 \over 2(\nu +
1)}} \ \
 \ .
\eqno(43)$$

\noindent
It is easily seen (see Appendix) that the factor $[(\nu +
3)/2(\nu+2)]^{1\over 2(\nu+1)}$ is an
increasing function of $\nu$ and goes to 1 as $\nu \rightarrow \infty$.

On the other hand the function ${\pi \over 2}{{1 \over \sqrt{\nu+2}}}
\left[({\nu+3 \over
2(\nu+2)}\right]^{{1 \over 2(\nu+2)}}$ is a decreasing function of $\nu$
(see Appendix).  Therefore
this example shows that our constant $C_\nu$ is not too bad, and that it
cannot be improved more
than by a factor $[(\nu_C+3)/2(\nu_C+2)]^{{1 \over 2(\nu_C+1)}} = \left[
{1 \over 2} + {2 \over
\pi^2} \right]^{{2 \over \pi^2-4}} \simeq 0.887$.
\par

We should note here that the Schr\"odinger equation with the potential
(40) can
 be solved
exactly at zero energy. The regular solution is $\varphi = \sqrt{r} J_{L
+ {1
 \over 2}}
(\sqrt{\lambda} Z)$, where $Z = r^{1 + {\nu \over 2}}/\left ( 1 + {\nu
\over 2}
 \right
)$, and $L = - \nu /2(2 + \nu)$. The exact value of $\lambda$ for which
the
 first bound
state appears is given by $\varphi '(r = 1) =0$. This is a
transcendental
 equation whose
solution is not simple, and needs numerical computation. In principle,
this
 exact value
of $\lambda$ could be used instead of $\lambda_1$ of the Calogero's
sufficient
 condition.

\vskip 5 mm
\noindent \underbar{\bf Acknowledgement} \par
Two of us (K. C. and A. M.) would like to thank Masud Chaichian for warm
 hospitality at the
Theoretical Physics Institute of the University of Helsinki where this
work
 began.

\vfill \supereject \centerline{\bf \underbar{Appendix}} \par \vskip 5 mm
We must show that the derivative of $F(\nu ) = [(\nu + 3)/( \nu +
2)]^{1/2(\nu +
 1)}$ is
positive, which amounts to the same thing for the derivative of $G(\nu )
= \log
 F(\nu )$.
Now, if we write $H(\nu ) = 2 (\nu + 1)^2 G'(\nu )$, we have
$$H(\nu ) = {1 \over \nu + 2} - {2 \over \nu + 3} - \log {\nu + 3 \over
2( \nu +
 2)}
\eqno(A.1)$$

\noindent and, therefore,
$$H'(\nu ) = (\nu + 1) \left [ {1 \over ( \nu + 2)^2} - {1 \over ( \nu +
3)^2}
 \right ] > 0
\ \ \ . \eqno(A.2)$$

\noindent It follows that $H(\nu )$ is increasing for $\nu \geq 0$. Now,
$$H(0) \equiv 2 G'(0) = \log \left ( {4 \over 3} \right ) - {1 \over 6}
> 0 \ \
\ . \eqno(A.3)$$

\noindent Therefore, $H(\nu )$ is positive for $\nu \geq 0$, and the
same is
 true for
$G'(\nu )$. $\sq$

\par
Similarly, we consider the derivative of $\tilde G(\nu) = -{1 \over 2}
\log (\nu+2) +G (\nu)$.  If
we write $\tilde H(\nu) = 2(\nu + 1)^2 \tilde G^\prime(\nu)$ we find
$$
\tilde H(\nu) = - {(\nu+1)^2 \over \nu+2} + H(\nu)
\eqno(A.4)
$$
and, therefore
$$
\matrix{
\tilde H^\prime(\nu) &=& -{(\nu+1)(\nu+3) \over (\nu+2)^2} +
H^\prime(\nu)\cr
&=& (\nu+1) \left[ - {1 \over \nu+2} - {1 \over(\nu+3)^2} \right] < 0}
\eqno(A.5)
$$
Since $\tilde H(0) = \log {4 \over 3} - {2 \over 3} < 0$ it follows that
$\tilde H(\nu) < 0$ for all
$\nu \geq 0$.

\vfill \supereject \centerline{\bf \underbar{References}} \par \vskip 5
mm
 \item{1.} F.
Calogero, Comm. Math. Phys. \underbar{1}, 80 (1965).   \item{2.} J. H.
E. Cohn,
 J. London
Math. Soc. \underbar{40}, 523 (1965).   \item{3.} V. Bargmann, Proc.
Nat. Acad.
 Sci. U.S.A.
\underbar{38}, 961 (1952).  \item{4.} R. Jost and A. Pais, Phys. Rev.
 \underbar{82}, 840
(1951).  \item{5.} K. Chadan, A. Martin and J. Stubbe, J. Math. Phys.
 \underbar{36}, 1616
(1995).  \item{6.} J. Schwinger, Proc. Nat. Acad. Sci. U.S.A.
\underbar{47}, 122
 (1961).
\item{7.} V. Glaser, H. Grosse, A. Martin and W. Thirring, in E. Lieb,
B. Simon
 and A.
Wightman, (eds), Studies in Mathematical Physics, Princeton University
Press
 1976, pp.
169-194.
\item{8.} K. Chadan, A. Martin and J. Stubbe, Letters in Math. Physics,
to
 appear (1995).
\item{9.} K. Chadan and R. Kobayashi, C. R. Ac. Sci. Paris,
\underbar{IIb, 320},
 339
(1995).
\item{10.} K. Chadan, Nuovo Cimento \underbar{A58}, 191 (1968).
\item{11.} W. M. Frank, J. Math. Phys. \underbar{8}, 466 (1967).
\item{12} M. Reed and B. Simon, Methods of Modern Mathematical Physics
II, Academic Press, New York,
1975.
 \item{13.} F. Calogero, J. Math. Phys. \underbar{6}, 161 (1965).

 \bye

 \bye